\documentclass[aps,pra,twocolumn,letterpaper,superscriptaddress,10pt]{revtex4-1} 
\usepackage{amssymb,amsthm,amsmath,amsfonts}
\usepackage{graphicx,mathptmx}
\usepackage{upgreek}
\usepackage{textcomp}
\usepackage{amsmath}
\usepackage{array, multirow, bigdelim, makecell, booktabs}
\usepackage{color}
\usepackage{soul}

\usepackage[shortlabels]{enumitem}
\usepackage[pdftex,dvipsnames,usenames]{xcolor}
\usepackage[colorlinks=true,urlcolor=blue,citecolor=blue,linkcolor=blue]{hyperref}
\usepackage{gensymb}

\usepackage[dvipsnames]{xcolor}
\usepackage[normalem]{ulem}

\begin{document}

\title{Certifying position-momentum entanglement at telecommunication wavelengths}

\author{Lukas Achatz}
	\email{lukas.achatz@oeaw.ac.at}
	\affiliation{Institute for Quantum Optics and Quantum Information - IQOQI Vienna, Austrian Academy of Sciences, Boltzmanngasse 3, 1090 Vienna, Austria}
	\affiliation{Vienna Center for Quantum Science and Technology (VCQ), Vienna, Austria}
\author{Evelyn A. Ortega}
	\affiliation{Institute for Quantum Optics and Quantum Information - IQOQI Vienna, Austrian Academy of Sciences, Boltzmanngasse 3, 1090 Vienna, Austria}
	\affiliation{Vienna Center for Quantum Science and Technology (VCQ), Vienna, Austria}
\author{Krishna Dovzhik}
	\affiliation{Institute for Quantum Optics and Quantum Information - IQOQI Vienna, Austrian Academy of Sciences, Boltzmanngasse 3, 1090 Vienna, Austria}
	\affiliation{Vienna Center for Quantum Science and Technology (VCQ), Vienna, Austria}
\author{Rodrigo F. Shiozaki}
    \affiliation{Institute for Quantum Optics and Quantum Information - IQOQI Vienna, Austrian Academy of Sciences, Boltzmanngasse 3, 1090 Vienna, Austria}
    \affiliation{Departamento de F\'isica, Universidade Federal de S\~{a}o Carlos, Rodovia Washington Lu\'is, km 235—SP-310, 13565-905 S\~{a}o Carlos, SP,Brazil}
\author{Jorge Fuenzalida}
    \affiliation{Institute for Quantum Optics and Quantum Information - IQOQI Vienna, Austrian Academy of Sciences, Boltzmanngasse 3, 1090 Vienna, Austria}
	\affiliation{Vienna Center for Quantum Science and Technology (VCQ), Vienna, Austria}
	\affiliation{Fraunhofer Institute for Applied Optics and Precision Engineering IOF, Albert-Einstein-Str. 7, 07745 Jena, Germany}
\author{Sören Wengerowsky}
    \affiliation{Institute for Quantum Optics and Quantum Information - IQOQI Vienna, Austrian Academy of Sciences, Boltzmanngasse 3, 1090 Vienna, Austria}
	\affiliation{Vienna Center for Quantum Science and Technology (VCQ), Vienna, Austria}
	\affiliation{Current address: ICFO-Institut  de  Ciencies  Fotoniques,  The  Barcelona  Institute  of Science  and  Technology,  08860  Castelldefels  (Barcelona),  Spain}
\author{Martin Bohmann}
    \email{martin.bohmann@oeaw.ac.at}
    \affiliation{Institute for Quantum Optics and Quantum Information - IQOQI Vienna, Austrian Academy of Sciences, Boltzmanngasse 3, 1090 Vienna, Austria}
	\affiliation{Vienna Center for Quantum Science and Technology (VCQ), Vienna, Austria}
\author{Rupert Ursin}
    \email{rupert.ursin@oeaw.ac.at}
	\affiliation{Institute for Quantum Optics and Quantum Information - IQOQI Vienna, Austrian Academy of Sciences, Boltzmanngasse 3, 1090 Vienna, Austria}
	\affiliation{Vienna Center for Quantum Science and Technology (VCQ), Vienna, Austria}

\begin{abstract}
    The successful employment of high-dimensional quantum correlations and its integration in telecommunication infrastructures is vital in cutting-edge quantum technologies for increasing robustness and key generation rate.
    Position-momentum Einstein-Podolsky-Rosen (EPR) entanglement of photon pairs are a promising resource of such high-dimensional quantum correlations.
    Here, we experimentally certify EPR correlations of photon pairs generated by spontaneous parametric down-conversion (SPDC) in a nonlinear crystal with type-0 phase-matching at telecommunication wavelength for the first time.
    To experimentally observe EPR entanglement, we perform scanning measurements in the near- and far-field planes of the signal and idler modes.
    We certify EPR correlations with high statistical significance of up to 45 standard deviations.
    Furthermore, we determine the entanglement of formation of our source to be greater than one, indicating a dimensionality of greater than 2.
    Operating at telecommunication wavelengths around $1550\,$ nm, our source is compatible with today's deployed telecommunication infrastructure, thus paving the way for integrating sources of high-dimensional entanglement into quantum-communication infrastructures.
\end{abstract}

\date{\today}
\maketitle

\section{Introduction}

    Recent scientific research in quantum information and quantum technologies focuses more and more on the exploitation of high-dimensional quantum correlations and its efficient integration into telecommunication infrastructures. 
    In particular, high-dimensional photonic entanglement in various degrees of freedom \cite{mirhosseini2015,gerke2015,steinlechner2017,wang2018,ansari2018,kysela2020,hu2020,chen2020,erhard2020} are an essential resource for quantum information applications. 
    Such high-dimensional systems are particularly interesting for quantum information applications as they can be used to encode several qubits per transmitted information carrier \cite{tittel2000,aolita2007,ali-khan2007,mower2013,graham2015}.
    Furthermore, high-dimensional entanglement features robustness against noise \cite{ecker2020,doda2020,hu2020b} that makes it an ideal candidate for realistic noisy and lossy quantum communication.

    One physical realization for high-dimensional entanglement are continuous-variable position-momentum correlations of photon pairs emitted by spontaneous parametric down-conversion (SPDC) \cite{walborn2010}.
    Because of energy and momentum conservation in the SPDC process, the produced continuous-variable states feature Einstein, Podolsky, and Rosen (EPR) correlations \cite{EPR} that can be certified via suitable witness conditions  \cite{reid,reid1989,reid2009, cavalcanti2009}.
    Importantly, the certification of EPR-correlations directly implies entanglement.
    Ideally, the entangled photon pairs manifest perfect position correlations or perfect transverse momentum anti-correlations, depending on the respective basis choice, which is implemented through different lens configurations.
    Thus, measurements of the two conjugate variables, position and momentum, of both photons allows to experimentally determine EPR entanglement.
    It is worth mentioning that this form of EPR-correlations in terms of position and momentum variables coincides with the original idea by EPR \cite{EPR}.
    The entanglement dimensionality \cite{terhal2000} of the signal and idler photons can, for example, be characterized by the entanglement of formation $E_F$ \cite{bennett1996}.
    
    The first experimental verification of EPR position-momentum entanglement of photons has been reported in \cite{howell2004}.
    Typically, sources of position-momentum entangled photons are based on type-I and type-II phase-matched SPDC sources; see, e.g., \cite{hale2005, ostermeyer2009, howell2004, moreau2014, edgar2012}.
    In these implementations, either scanning techniques \cite{howell2004, ostermeyer2009, hale2005, schneeloch2016} or cameras \cite{edgar2012, eckmann2020} were used to record the position and momentum observables by means of near- and far-field measurements, respectively.
    High $E_F$ values and thus high-dimensional position-momentum entanglement of photons have been reported in \cite{schneeloch2019}.
    Applications of position-momentum entangled photons are discussed in the literature, which includes ideas for quantum key distribution (QKD) \cite{almeida2005_I}, continuous-variable quantum computation \cite{tasca2011}, ghost-imaging applications \cite{pittmann1995} and dense-coding \cite{bennett1992}.

    Despite its huge potential as mentioned above, high-dimensional position-momentum EPR-entanglement from SPDC sources are underrepresented among the quantum information applications.
    This has practical reasons. 
    Firstly, the direct detection of photons using cameras makes further manipulations of the photons impossible and thus hinders the application of quantum information protocols.
    Secondly, the wavelength of the photons used in the majority of position-momentum experiments is not compatible with the current telecommunication infrastructure which works best between 1260 nm and 1625 nm, where optical loss is lowest \cite{ghatak2000}.
    Additionally, most experiments use type-II phase-matched SPDC-crystals to produce the entangled photon pairs, which have a relatively low brightness and narrow spectrum as compared to type-0 phase-matched SPDC-crystals \cite{steinlechner2014}.
    
    In this paper, we report on a bright source producing position-momentum EPR-entangled photons at telecommunication wavelengths allowing for the efficient integration into existing quantum communication infrastructures.
    We generate photons via SPDC in a Magnesium Oxide doped periodically poled Lithium Niobate (MgO:ppLN) crystal with type-0 phase-matching in a Sagnac-configuration with a center-wavelength of $1550.15\,$ nm.
    We evaluate near- and far-field correlations by probabilistically splitting the photon-pairs at a beamsplitter and transversely scanning two fibers in ($x\textsubscript{1}, y\textsubscript{1}$) and ($x\textsubscript{2}, y\textsubscript{2}$) directions respectively.
    The photons in each fiber were counted with a superconducting nanowire single-photon detector (SNSPD) and a time tagging device connected to a readout-system.
    The temporal correlation between the time tags was later analyzed. 
    Based on these measurements, we certify position-momentum EPR-entanglement with high statistical significance.
    We further calculate the entanglement of formation of the photon pairs to be greater than 1, indicating a dimensionality of greater than 2.
    Furthermore, our setup is stable over several hours and the used imaging system is capable of coupling the entangled photons into photonic waveguides.
    In addition, our source can be extended to hyper-entanglement experiments harnessing entanglement in the position-momentum as well as polarization degree of freedom.
    To the best of our knowledge, this is the first demonstration of position-momentum EPR-entangled photon pairs from a type-0 phase-matched SPDC source at telecommunication wavelength.
    Our results show the possibility of utilizing the high-dimensional characteristic of continuous-variable EPR-entanglement for this regime and, thus, enable its integration in existing telecommunication quantum-communication systems.

\section{Experiment}

    \begin{figure}
    \includegraphics[width=0.45\textwidth]{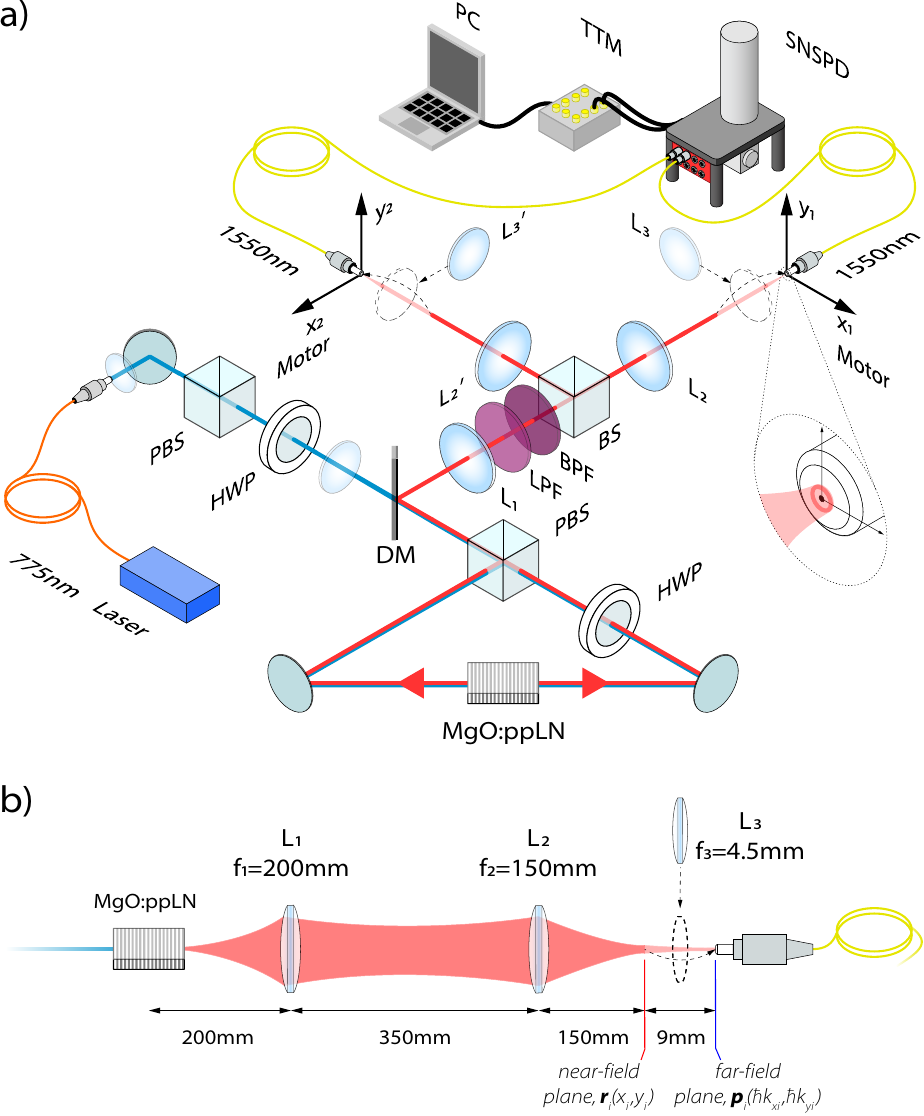} 
        \caption{\textbf{a)} Setup used to measure the coincidences in near- and far-field. The laser emits light at a center wavelength of $775.078\,$ nm. The beam is slightly focused into a 40-mm-long ppLN type-0 phase-matched crystal. After the crystal two lens configurations are used to implement the near- and far-field measurements. The focal lengths of the lenses are $f_1$ = 250 mm, $f_2$ = $f_2'$ = 200 mm and $f_3$ = $f_3'$ = 4.5 mm.
        \textbf{b)} Simplified setup showing the lens configurations. In analogy with the polarization basis, different lens configurations correspond to different complementary basis choices. 
        For the near-field measurement (position-space, photon pairs show correlation), $\text{L}_3$ ($\text{L}_3'$) was not mounted in the setup and the fiber was placed in the focal point of lens $\text{L}_2$ ($\text{L}_2'$).
        For the far-field measurement (momentum-space, photon pairs show anti-correlation), $\text{L}_3$ ($\text{L}_3'$) was mounted at the focal distance $f_2 + f_3$ ($f_2' + f_3'$) and the fiber was moved back for 9 mm, thus aligning it with the focal distance of $\text{L}_3$ ($\text{L}_3'$).
        }
        \label{fig:setup_far_near_field} 
    \end{figure}

    In our experiment, we studied and characterized the position-momentum entanglement of photon pairs.
    For this purpose, we generated photon pairs at telecommunication wavelength from a type-0 phase-matched SPDC-source in a Sagnac configuration \cite{kim2006}.
    By measuring the distributions of the signal and idler photons in the near- and far-field we obtained their position and momentum correlations.
    With these we can investigate EPR-correlations in position and moment of the photon pairs.
    It is worth mentioning that certifying EPR-correlations directly implies entanglement between signal and idler photons.
    In this sense, demonstrating EPR-correlations is a stronger and more direct indication of the non-local character of quantum mechanics \cite{reid2009} which, however, is in general more demanding than verifying entanglement.
    
    The experimental setup is shown in Fig. \ref{fig:setup_far_near_field}.
    The heart of the SPDC-source is a $40$-mm-long ppLN bulk crystal with a poling period of $19.2$ $\upmu$m.
    The crystal was pumped with a continuous-wave (CW) laser at $775.078\,$ nm, $10\,$ mW pump power and a pump-beam waist (FWHM) at the center of the crystal of $83.5\,$ $\upmu$m.
    In the parametric process, one pump photon is converted with low probability to a pair of signal and idler photons in the telecommunication wavelength regime of around $1550\,$ nm.
    The emission angles of the emitted photons during the SPDC process strongly depend on the temperature of the crystal.
    By controlling the temperature of the crystal we adjust a non-collinear, quasi-phase-matched SPDC process; see \cite{rodrigo2020} for details.
    Due to the momentum conservation in the SPDC process, the signal and idler photons are emitted with opposite transverse ($x$-$y$-plane) momentum-vectors, meaning that $\textbf{p}_1= -\textbf{p}_2$ with $\textbf{p}_i=(\hbar k_{x_i},\hbar k_{y_i})$, where $k$ refers to the transverse wave-vector and $i=1,2$. 
    Thus, when imaging the photons in the far-field plane of the crystal, one can observe anti-correlations in the coincidence counts.
    From hereon indices 1 and 2 refer to signal and idler photon, respectively. 
    Since the photons of each pair are created at the same point in space, imaging the photons in the near-field results in correlations, meaning that $\textbf{r}_1 = \textbf{r}_2$, with $\textbf{r}_i=(x_i,y_i)$ and $i=1,2$.
    
    Note that through the Sagnac configuration the nonlinear crystal is bidirectionally pumped.
    In this way, the produced photon pairs not only exhibit position-momentum entanglement but are also entangled in polarization \cite{kim2006}, although this degree of freedom was not investigated in this work.

    After the Sagnac loop, the SPDC photons were separated from the pump beam with a dichroic mirror (DM) and spectrally cleaned with a long-pass filter (LPF, $\lambda_{\mathrm{cut-off}} = 780\,$ nm) and a band-pass filter (BPF, $\lambda_{\mathrm{BPF}} = 1550\pm3\,$ nm).
    After the first lens ($\text{L}_1$), a beam splitter (BS) separates signal and idler photons probabilistically for $50\%$ of the incoming pairs.
    To perform the near-field (position) and far-field (momentum) measurements of the photon pairs, two different lens configurations are used; cf. also Fig. \ref{fig:setup_far_near_field} b).
    The far-field measurement of the signal and idler photons is implemented via $\text{L}_1$ with a focal length of $f_1 = 200\,$ mm. $\text{L}_1$ acts as a Fourier lens that maps the transverse position of the photons to transverse momentum in the far-field plane.
    Then, the far-field plane is imaged and demagnified by two consecutive lenses $\text{L}_2$ ($\text{L}_2'$) and $\text{L}_3$ ($\text{L}_3'$) with focal lengths of $f_2 = f_2' = 150\,$ mm and $f_3 = f_3' = 4.5\,$ mm, resulting in a demagnification factor of $M_{\mathrm{FF}}=f_3/f_2 = f_3'/f_2' =0.03$.
    This results in an effective focal length of the Fourier lens of $6\,$mm.
    For the near-field measurement, the crystal plane is imaged with a $2f/2f'$ optical system by $\text{L}_1$ and $\text{L}_2$ ($\text{L}_2'$).
    This configuration results in a magnification factor of $M_{\mathrm{NF}}= f_2/f_1 = f_2'/f_1 = 0.75$.
    Note that the used imaging system would allow for the integration of the near and far-field measurements in (micro-)photonic architectures, such as multicore fibers \cite{Xavier2020} or photonic chips \cite{wang2020}.

    In two consecutive measurement runs, we measured the coincidences of the correlated photons in the near- and far-field.
    A coincidence count was identified when both detector-channels register photons within a $300\,$ ps time-window.
    These measurements were achieved by performing a transverse plane-scan with two single-mode fibers (SMF), which guide the photons to a SNSPD and read-out electronics. 
    Such scanning approaches have already been used to characterize the spatial correlations for type-II  SPDC-sources \cite{ostermeyer2009, howell2004, hale2005}.
    Arrival times were tagged using a time-tagging-module (TTM), combining the photon counts with the positions of the translation stages at all times.
    The scans were performed for both settings (near- and far-field) in the signal and idler arms.
    Using motorized translation stages, each coordinate was sampled in $17$ steps with a step size of $10$ $\upmu$m, which results in a $17\times17$ grid on each detection side and a total of $17^4=83.521$ data points.
    For each data point we measured $1$ s and recorded the single counts of each detector and the coincidence-counts between both detectors.
    The total measurement time was about $80\,$ h, throughout which the whole setup was stable.
\section{Results}

\subsection{Scan results}
\begin{figure}
        \includegraphics[width=1\columnwidth]{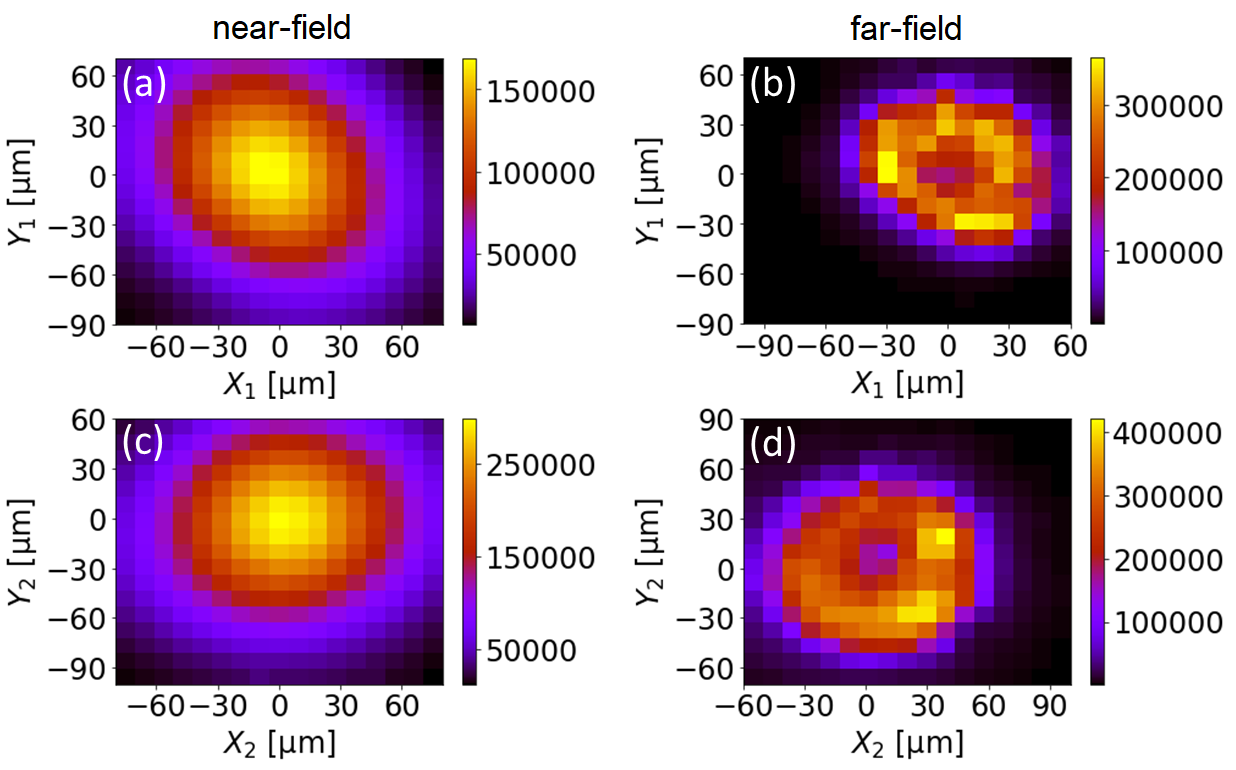}
        \caption{\label{fig:singles_near_far_field}
        Measured average single counts per second for every $(x_i, y_i)$ in the near-field (a),(c) and the far-field (b),(d) for the signal $(i=1)$ and idler $(i=2)$ modes.
        To obtain the average single counts per second for every $(x_1, y_1)$ position, we averaged the recorded counts of the signal mode over all positions of the $(x_2, y_2)$ detection stage, which corresponds to an integration time of $17^2$ s.
        The same procedure is performed to determine the average single counts of the idler beam along $(x_2, y_2)$.
        }
    \end{figure}

\begin{figure}
        \includegraphics[width=1\columnwidth]{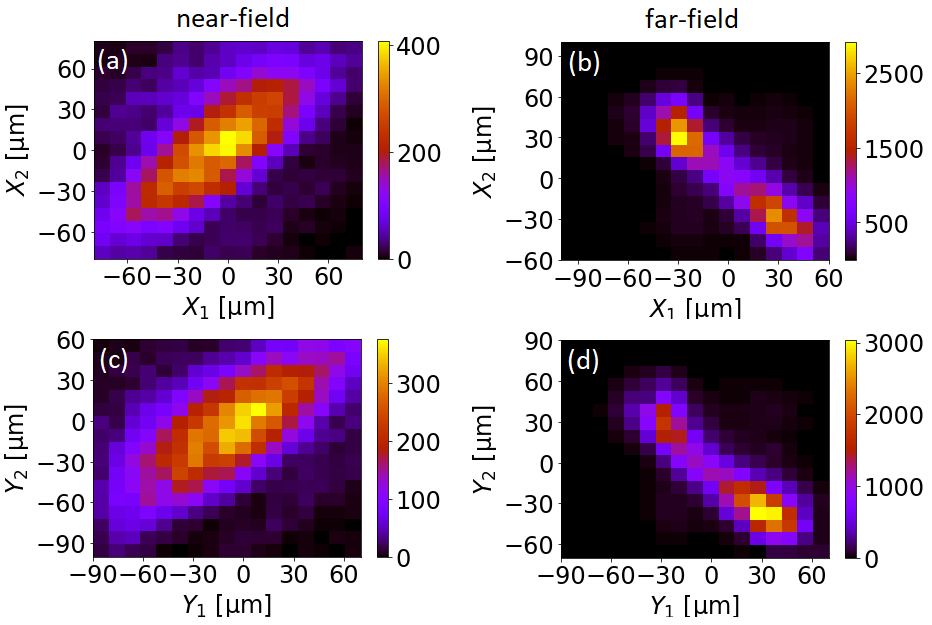}
        \caption{\label{fig:cc_near_far_field} 
        Measured coincident counts for every $(x_1, x_2)$ and $(y_1, y_2)$ in the near-field (a),(c) and the far-field (b),(d). For plotting the $(x_1, x_2)$ case, the $(y_1, y_2)$ fiber positions were taken to be at 0 $\upmu$m. For the $(y_1, y_2)$ case, the $(x_1, x_2)$ fiber positions were taken to be at 0 $\upmu$m.
        }
    \end{figure}

    The results for the near- and far-field scans are summarized in Figs. \ref{fig:singles_near_far_field} and \ref{fig:cc_near_far_field}.
    For both measurement settings (either near- or far-field), we obtained the single counts at the motor-positions $(x_1, y_1)$ and $(x_2, y_2)$ as well as the coincidence counts between both detection stages at $(x_1, y_1, x_2, y_2)$.
    The maximum single-photon count-rate in the near-field (far-field) was about 150 kHz (300 kHz) for signal and about 250 kHz (400 kHz) for the idler mode. 
    The maximum coincident-photon count-rate in the near-field (far-field) was about 500 Hz (3000 Hz) for both signal and idler modes.
    The discrepancy between the count rates of signal and idler modes can be explained by different coupling efficiencies into the single-mode fibers.
    To obtain the average single counts for every $(x_1, y_1)$ position, we averaged the recorded counts of the signal mode over all positions of the $(x_2, y_2)$ detection stage, which corresponds to an integration time of $17^2$ s.
    The same procedure is performed to determine the average single counts of the idler beam along $(x_2, y_2)$.
    Figure \ref{fig:singles_near_far_field} displays the averaged single counts in the signal and idler modes.
    In Fig. \ref{fig:cc_near_far_field} the correlation between every $(x_1, x_2)$ and $(y_1, y_2)$ coordinate pair is shown.
    
    In both figures, the expected features of the non-collinear SPDC emission in the near- and far-field are reflected: in the near-field, the photon pairs follow a Gaussian distribution.
    For the single counts the shape is basically given by the intensity distribution (beam waist) of the pump beam in the center of the nonlinear crystal, while for the coincident counts the Gaussian form is given by the form of the joint wave function in \textbf{r}-space \cite{monken1998}. 
    And in far-field the anti-correlations in the transverse momenta, stemming from the non-collinear quasi-phase-matched SPDC process, manifests itself in a doughnut or ring-shape structure.
    We note that no data-correction or post-processing, such as subtraction of dark counts, was performed.
    In particular, no accidental coincidences were subtracted as is required in experiments using cameras \cite{edgar2012, eckmann2020}.
    The visible asymmetries can be explained by imperfect optical components and changing focusing parameters during the scans.
    
    When comparing the single counts of signal and idler modes, a discrepancy of about 30\% is visible.
    This can be explained by the different coupling efficiencies of the used fibers.
    The slight difference in the count rates does, however, not influence the quality of the measured entanglement, as it is analyzed based on coincidence detection.
    
    Furthermore, from Fig. \ref{fig:singles_near_far_field} (a) and (c) and taking into account the magnification of the lens-setup we estimate the diameter of the photon-pair birth region inside the crystal to be $\approx 80$ $\upmu$m.
    This is in very good agreement with the pump-beam waist of $83.5$  $\upmu$m at the crystal and thus provides a suitable sanity check of the results.
    
    Based on the recorded scans in the near- and far-field of the signal and idler photons, we now analyze the position and momentum quantum correlations of our telecom-wavelength type-0 phase-matched SPDC-source.
    Firstly, we will verify the EPR-type entanglement based on the spatial spread between the coincidence counts of signal and idler photons.
    Secondly, we estimate the entanglement dimensionality by determining $E_F$ in both scan directions.

\subsection{Certifying EPR entanglement}

    \begin{figure}
        \includegraphics[width=1\columnwidth]{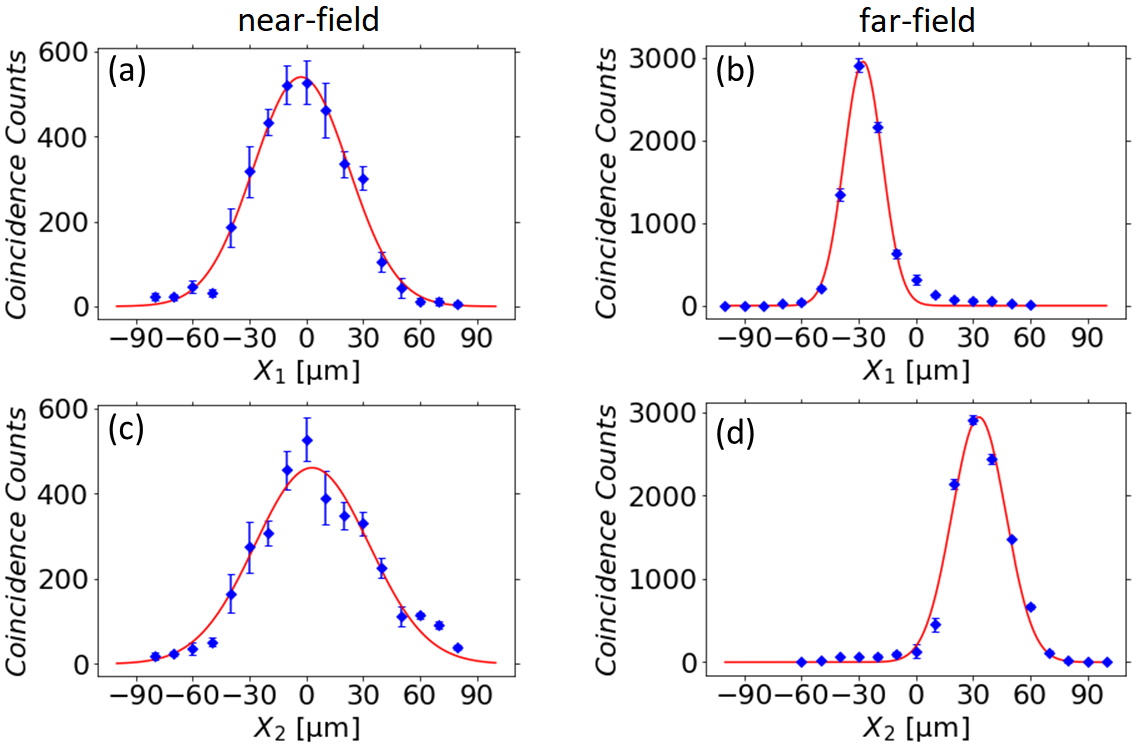}
        \caption{\label{fig:cc_near_far_field_gauss_x}
        Example cases for the Gauss-fits of the obtained coincidence counts for the x coordinate. For calculating the minimum inferred variance for the entire $x_1$-axis according to Eq. \eqref{eq:variance}, the fiber positions of $x_2$ were ranging from -40 $\upmu$m to +40 $\upmu$m with a stepsize of 10 $\upmu$m. At every step the data was fitted and the standard deviation calculated. The values on the abscissa represent the position of the fibers. The minimum inferred variances were then obtained by integration over the respective coordinate, according to Eq. \eqref{eq:variance}.
        \textbf{(a)} Coincidence counts in the near-field for $(x_1, y_1, x_2, y_2) = (x_1, 0, 0, 0)$.
        \textbf{(b)} Coincidence counts in the far-field for $(x_1, y_1, x_2, y_2) = (x_1, 0, +30$ $\upmu \text{m}, 0)$.
        \textbf{(c)} Coincidence counts in the near-field for $(x_1, y_1, x_2, y_2) = (0, 0, x_2, 0)$.
        \textbf{(d)} Coincidence counts in the far-field for $(x_1, y_1, x_2, y_2) = (-30$ $\upmu \text{m}, 0,x_2, 0)$.
        }
    \end{figure}
    
   Position-momentum entanglement of the signal and idler modes can be certified by violating a so-called EPR-Reid criterion. For this we are introducing the following local-realistic premises \cite{EPR, reid2009}:
    \\
    Realism: \textit{If, without in any way disturbing a system, we can predict with some specified uncertainty the value of a physical quantity, then there exists a stochastic element of physical reality which determines this physical quantity with at most that specific uncertainty.} \\
    Locality: \textit{A measurement performed at a spatially separated location 1 cannot change the outcome of a measurement performed at a location 2.}\\
    
    In the following argument the receiver of the signal photon is referred to as Alice, while the receiver of the idler photon is referred to as Bob. The argumentation goes as follows: When Bob measures the position $x_2$ of particle 2 he can then estimate, with some uncertainty, the position $x_1$ of particle 1 at Alice's site. The average inference variance of this estimate $x_1^{\text{est}}(x_2)$ is defined as follows \cite{cavalcanti2009, reid2009}: 
    \begin{equation}
    \Delta_{\mathrm{inf}}^2(x_1|x_2) \equiv \int dx_1 P(x_1|x_2) \left[x_1 - x_1^{est}(x_2)\right]^2
    \end{equation}
    where $P(x_1|x_2)$ is the conditional probability to measure $x_1$ if $x_2$ has already been measured. 
    The minimum of this inferred variance is obtained when the estimated value $x_1^{\text{est}}(x_2)$ is the expectation value of $x_1$. Thus, the minimum inferred variance for a position measurement is given by
    \begin{equation}
    \Delta_{\mathrm{min}}^2(\textbf{r}_1|\textbf{r}_2) = \int d\textbf{r}_2 P(\textbf{r}_2)\Delta^2(\textbf{r}_1|\textbf{r}_2)
    \label{eq:variance}
    \end{equation}
    and
    \begin{equation}
    \Delta_{\mathrm{min}}^2(\textbf{p}_1|\textbf{p}_2) = \int d\textbf{p}_2 P(\textbf{p}_2)\Delta^2(\textbf{p}_1|\textbf{p}_2)
    \label{eq:variance}
    \end{equation} 
    for a momentum measurement. Here $\Delta^2(\textbf{r}_1|\textbf{r}_2)$ is the uncertainty (i.e. variance) in $\textbf{r}_1$ for a fixed $\textbf{r}_2$ and $P(\textbf{r}_2)$ represents the probability distribution of $\textbf{r}_2$ which are experimentally determined via the relative count frequencies. The minimum inferred variance for the transverse momenta $\Delta_{\mathrm{min}}^2(\textbf{p}_1|\textbf{p}_2)$ is defined analogously.
    The argument continues as follows: Since Bob is able to infer the outcome of either a position or momentum measurement at Alice's site within some uncertainty and since the locality assumption does not allow a measurement to induce change at a spatially separated location, in a local-realistic picture it follows that statistical elements of reality determining the position and momentum of both particles must exist at the same time.
    The precision with which one can measure the position and momentum of a particle is fundamentally limited by Heisenberg's uncertainty principle $\Delta x \Delta p \geq \frac{\hbar}{2}$. By measuring the position and momentum of particles with higher accuracy than permitted by Heisenberg's uncertainty principle, the local-realistic assumptions cannot hold.
    Thus, EPR-correlations can be certified by violating following inequality \cite{reid1989,cavalcanti2009,reid2009}:
    \begin{equation}
        \Delta_{\mathrm{min}}^2(\textbf{r}_1|\textbf{r}_2) \Delta_{\mathrm{min}}^2(\textbf{p}_1|\textbf{p}_2) > \frac{\hbar^2}{4}.
        \label{eq:inequality}
    \end{equation}
    Here, $\Delta_{\mathrm{min}}^2(\textbf{r}_1|\textbf{r}_2)$ is the minimum inferred variance, which represents the minimum uncertainty in inferring transverse position $\textbf{r}_1$ of the signal mode conditioned upon measuring $\textbf{r}_2$ in the idler mode.
     
    For the calculation of the minimum inferred variances in the x-direction, we only took into account the values on the abscissa, meaning that $y_1 = y_2 = 0$. 
    As for the y-direction, only values along the ordinate axis, where $x_1 = x_2 = 0$, were taken into account.
    We also treated the $x$ and $y$ components separately as they correspond to commuting observables.

    In accordance with previous experiments \cite{hale2005, ostermeyer2009, howell2004, moreau2014, edgar2012}, we determined the minimal inferred variances in near and far-field by fitting the conditional uncertainties $\Delta^2(\textbf{r}_1|\textbf{r}_2)$ and $\Delta^2(\textbf{p}_1|\textbf{p}_2)$ with  Gaussian functions of the form (here formulated for $x_1$ with a fixed position $x_2$):
    \begin{equation}
        \Psi (x_1|x_2) = \mathcal{N} \text{exp}\left[\frac{-(x_1-\mu)^2}{4\sigma^2_{x_{1}}}\right]
        \label{eq:gaussian}
    \end{equation}
    where $\mathcal{N}$ is a normalization constant, $\mu$ is the expectation value and $\sigma_{x_{1}}$ is the standard deviation, i. e. the uncertainty along the $x_1$ direction.
    For an extensive discussion of this approach the reader is referred to \cite{Schneeloch_2016_2}.
    Exemplarily, we show for four cases the data and the corresponding Gaussian fits in Figure \ref{fig:cc_near_far_field_gauss_x}.
    From the obtained standard deviations one can then calculate the conditional near-field variances by taking into account the magnification of our imaging system
    \begin{equation}
        \Delta^2(x\textsubscript{1}|x\textsubscript{2}) = \left( \frac{\sigma_{x_{1}}}{M_{\mathrm{NF}}}\right)^2.
        \label{eq:sd_near_field}
    \end{equation} 
    The same procedure is repeated for $y_1$, $x_2$ and $y_2$.
    
    For the far-field measurement the procedure is analogous.
    To translate the obtained standard deviations to the actual conditional momentum variances following relation is used \cite{saleh1991}:
    \begin{equation}
        \Delta^2(p\textsubscript{x\textsubscript{1}}|p\textsubscript{x\textsubscript{2}}) = \left( \frac{2\pi \hbar}{\lambda f_1 M_{\text{FF}}} \sigma_{{x_{1}}} \right)^2,
        \label{eq:sd_far_field}
    \end{equation}
    where $\lambda$ is the wavelength of the entangled photons, $f_1$ is the focal length of the Fourier lens $\text{L}_1$ and $\sigma_{{x_{1}}}$ is the standard deviation as obtained by the Gauss-fit.
    The same procedure is repeated for $p\textsubscript{y\textsubscript{1}}$, $p\textsubscript{x\textsubscript{2}}$ and $p\textsubscript{y\textsubscript{2}}$.
    The obtained values for the minimum inferred variances calculated with Eq. \eqref{eq:variance} are listed in Table \ref{tab:variances}.

    \begin{table}
    \begin{center}
        \renewcommand{\theadfont}{\normalsize\bfseries}
         \begin{tabular}{c|c|c} 
         \hline\hline
         \thead{Minimum\\variance} & \thead{Obtained\\values}
         & \thead{Uncertainty\\product}\\ [0.5ex] 
         \hline\hline
        
         \rule{0pt}{3ex} $\Delta_{\text{min}}^2(x\textsubscript{1}|x\textsubscript{2})$ &  (6.6 $\pm$ 0.6) $\times$ $10^{-4}$ $\text{mm}^2$ & 
         \rdelim\}{2}{*}[$(2.4 \pm 0.5)  \times 10^{-2}$$\hbar^2$]\\ [1ex]
         
         $\Delta_{\text{min}}^2(p\textsubscript{x\textsubscript{1}}|p\textsubscript{x\textsubscript{2}})$ &  (37.1 $\pm$ 6.1)$\hbar^2\text{mm}^{-2}$\\ [1ex]
         
         $\Delta_{\text{min}}^2(y\textsubscript{1}|y\textsubscript{2})$ & (9.2 $\pm$ 0.7) $\times$ $10^{-4}$ $\text{mm}^2$ 
         & \rdelim\}{2}{*}[$(3.7 \pm 0.6)  \times 10^{-2}$$\hbar^2$]\\ [1ex] 
         
         $\Delta_{\text{min}}^2(p\textsubscript{y\textsubscript{1}}|p\textsubscript{y\textsubscript{2}})$ &  (39.8 $\pm$ 6.3)$\hbar^2\text{mm}^{-2}$\\ [1ex]
         
         $\Delta_{\text{min}}^2(x\textsubscript{2}|x\textsubscript{1})$ &  (7.8 $\pm$ 0.9) $\times$ $10^{-4}$ $\text{mm}^2$ &
         \rdelim\}{2}{*}[$(3.7\pm 0.7)  \times 10^{-2}$$\hbar^2$]\\ [1ex]
         
         $\Delta_{\text{min}}^2(p\textsubscript{x\textsubscript{2}}|p\textsubscript{x\textsubscript{1}})$ &  (46.8 $\pm$ 6.8)$\hbar^2$$\text{mm}^{-2}$\\ [1ex]
         
         $\Delta_{\text{min}}^2(y\textsubscript{2}|y\textsubscript{1})$ & (7.0 $\pm$ 0.5) $\times$ $10^{-4}$ $\text{mm}^2$ &
         \rdelim\}{2}{*}[$(3.4 \pm 0.5)  \times 10^{-2}$$\hbar^2$]\\ [1ex] 
         
         $\Delta_{\text{min}}^2(p\textsubscript{y\textsubscript{2}}|p\textsubscript{y\textsubscript{1}})$ & (48.1 $\pm$ 6.9)$\hbar^2\text{mm}^{-2}$\\ [1ex]
         \hline\hline
        \end{tabular}
    \caption{Obtained minimum variances from fitting a Gaussian-function in the form of Eq. \eqref{eq:gaussian} to the obtained raw-data. The obtained values are calculated according to Eq. \eqref{eq:sd_near_field} for $\Delta_{\text{min}}^2(x\textsubscript{i}|x\textsubscript{j})$ and Eq. \eqref{eq:sd_far_field} for $\Delta_{\text{min}}^2(q\textsubscript{x\textsubscript{i}}|q\textsubscript{x\textsubscript{j}})$. The uncertainty product is then calculated by multiplying the minimum inferred variances. A clear violation of inequality \eqref{eq:inequality} is shown. The errors of the obtained values stem both form the Poisson distribution of the photons as well as the Gaussian fit, while the errors of the uncertainty product were calculated with standard Gaussian error propagation.}
    \label{tab:variances}
    \end{center}
\end{table}
    
    Based on the minimal inferred variances we can now test for EPR correlations via Eq. \eqref{eq:inequality}.
    The values of the left-hand-side of Eq. \eqref{eq:inequality} for the different cases are listed in Table \ref{tab:variances}.
    In all cases, we observe a clear violation of the inequality which certifies EPR-correlations between the signal and idler photons in their position-momentum degree of freedom.
    In particular, inequality \eqref{eq:inequality} is violated by up to 45 standard deviations, which demonstrates the high statistical significance of our results  as well as the degree to which our system deviates from a local-realistic one.
    Thus, we could certify position-momentum entanglement from a type-0 phase-matched SPDC-source at telecommunication wavelength.


\subsection{Entanglement of formation and entanglement dimensionality}
    
    A key advantage of continuous-variable entanglement lies in the fact that, in principle, the underlying infinite-dimensional Hilbert can be exploited \cite{horodecki2009}.
    This means that high-dimensional quantum information protocols can be employed which go beyond the entanglement structure of qubit systems as, for example, in the case of polarization entangled photon pairs.
    To certify high-dimensional entanglement in our experiment, we use
    the entanglement of formation $E_{\mathrm{F}}$ that is defined as the number of Bell states necessary to fully describe the system \cite{bennett1996}.
    Importantly, $E_{\mathrm{F}}$  provides a lower bound to the entanglement dimensionality $D$ through $D\geq 2^{E_{\mathrm{F}}}$.
    In particular, this means that $E_{\mathrm{F}}>1$ implies an entanglement dimensionality of at least 3.
    Hence, the entanglement of formation serves as a way of certifying and quantifying high-dimensional entanglement.

    To characterize the dimensionality of the position-momentum entanglement attainable from our source, we estimate the $E_{\mathrm{F}}$ of the entangled photon pairs along the $x$-axis via \cite{schneeloch2018}:
    \begin{equation}
        E_{\mathrm{F}} \geq -\text{log}_{2}\left(\text{e} \hspace{0.1cm} \Delta(x\textsubscript{1}|x\textsubscript{2}) \hspace{0.1cm} \Delta(k\textsubscript{x\textsubscript{1}}|k\textsubscript{x\textsubscript{2}})\right),
    \end{equation}
    where $\Delta(x\textsubscript{1}|x\textsubscript{2})$ and $\Delta(k\textsubscript{x\textsubscript{1}}|k\textsubscript{x\textsubscript{2}})$ are the measured standard deviations as obtained by the Gauss-fit from Eq. \eqref{eq:gaussian} and transformed with Eqs. \eqref{eq:sd_near_field} and \eqref{eq:sd_far_field}, respectively.
    For this purpose, we consider the scans where the count rates are highest, which is along $(x_1, y_1, x_2, y_2) = (x_1, 0, 0, 0)$ in the near-field plane and $(x_1, y_1, x_2, y_2) = (x_1, 0, +30$ $\upmu \text{m}, 0)$ in the far-field plane, stemming from the non-collinear SPDC process. 
    The entanglement of formation along the $y$-axis is defined similarly, and we determine it based on the near- and far-field plane scans along $(x_1, y_1, x_2, y_2) = (0, y_1, 0, 0)$ and $(x_1, y_1, x_2, y_2) = (0, y_1, 0, +30$ $\upmu \text{m})$, respectively.
    The calculated values for the $x$ and $y$ scans and the resulting values of $E_{\mathrm{F}}$ are listed in Table \ref{tab:EoF}.
    For both directions $E_{\mathrm{F}}$ is greater than 1, implying an entanglement dimensionality of at least 3 in either direction.
    
    At this point it has to be stressed that the entanglement dimensionality strongly depends on the length of the SPDC crystal. 
    A shorter crystal results in a higher degree of position-momentum entanglement \cite{Schneeloch_2016_2}.
    In other experiments \cite{edgar2012, ostermeyer2009, howell2004, schneeloch2019} the crystal length does not exceed $5\,$mm, as compared to the $40\,$mm-long crystal used in this experiment, which explains the relatively low values of the calculated entanglement of formation and entanglement dimensionality and also suggests that, in principle, a much higher entanglement dimensionality could be achieved with our setup by adapting the crystal length.

    \begin{table}
    \begin{center}
        \renewcommand{\theadfont}{\normalsize\bfseries}
         \begin{tabular}{c|c|c} 
         \hline\hline
         \thead{Standard\\deviation} & \thead{Obtained\\values}
         & \thead{$E_{\mathrm{F}}$}\\ [0.5ex] 
         \hline\hline
        
         \rule{0pt}{3ex} $\Delta(x\textsubscript{1}|x\textsubscript{2})$ &  (3.12 $\pm$ 0.15) $\times$ $10^{-2}$ $\text{mm}$ & 
         \rdelim\}{2}{*}[$(1.30 \pm 0.10)$]\\ [1ex]
         
         $\Delta(k\textsubscript{x\textsubscript{1}}|k\textsubscript{x\textsubscript{2}})$ &  (4.79 $\pm$ 0.24) $\text{mm}^{-1}$\\ [1ex]
         
         $\Delta(y\textsubscript{1}|y\textsubscript{2})$ & (3.18 $\pm$ 0.19) $\times$ $10^{-2}$ $\text{mm}$ 
         & \rdelim\}{2}{*}[$(1.18 \pm 0.13)$]\\ [1ex] 
         
         $\Delta(k\textsubscript{y\textsubscript{1}}|k\textsubscript{y\textsubscript{2}})$ &  (5.09 $\pm$ 0.35) $\text{mm}^{-1}$\\ [1ex]
        
         \hline\hline
        \end{tabular}
    \caption{
    Obtained values for the entanglement of formation ($E_{\mathrm{F}}$) based on the conditional standard deviations along the $x$- and $y$-axes scans in the near- and far-field.
    The standard deviations are obtained by fitting Gaussian functions in the form of Eq. \eqref{eq:gaussian} to the data and using Eqs. \eqref{eq:sd_near_field} and \eqref{eq:sd_far_field}.}
    \label{tab:EoF}
    \end{center}
\end{table}


    \section{Conclusion}

    Our experiment exhibits several important features which need to be highlighted.
    
    Firstly, it is important to stress that the entangled photon pairs are created at wavelengths around $1550\,$nm, thus at the C-Band of the telecommunication infrastructure.
    Operating sources in this wavelength regime allows for the integration into modern and future telecommunication infrastructures and increases the transmission rate over long distances.
    
    Secondly, with the possibility to distribute position-momentum-entangled photons efficiently, it is possible to exploit the entanglement for quantum communication purposes.
    This allows to implement quantum key distribution based on the photons' EPR correlations \cite{almeida2005_I}:
    by randomly choosing different measurement settings (near- and far-field) and distributing the photons to the two communication partners via two fibers a secret key can be shared.
    Recent developments in quantum communication using multicore fibers \cite{Xavier2020} also suggest that position-momentum-entangled photon pairs can be distributed through several cores simultaneously.
    Thus, multicore fibers provide a pathway to high-dimensional encoding and multiplexed distribution \cite{ortega2021}.
    In this context it is important to mention that the used imaging system facilitates the coupling of the entangled photon pairs into multicore fibers or other mircophotonic architectures, which allows for a direct coupling into such platforms.
    
    Thirdly, the presented experimental setup features the possibility to exploit quantum hyper-correlations, i.e., simultaneous entanglement in different degrees of freedom. 
    The Sagnac configuration of our source produces not only position-momentum entanglement but also entanglement between the polarization degrees of freedom of the photon pairs.
    Although we did not characterize the polarization entanglement in this experiment, it is possible to utilize this additional degree of freedom in hybrid quantum information tasks.
    Such a hybrid strategy can offer a significant increase of the overall dimensionality as it scales with the product of the dimensionalities of the individual degrees of freedom. 
    Finally, it is important to highlight the long-term stability of the experimental setup over a period of several days that renders the implementation of stable quantum information applications possible.

    The obtained standard deviations are in good agreement with values reported in previous experiments, as for example in \cite{howell2004, eckmann2020}, although about 2 orders of magnitude lower than the value reported in \cite{edgar2012}. When comparing the entanglement of formation with the latest publications, for example \cite{schneeloch2019, edgar2012}, it is apparent that the dimensionalities obtained here are far lower. The reason for this is the comparably long crystal and the focused pump beam. By adjusting and optimizing these parameters, the entanglement and hence the dimensionality can be increased.

    We have performed a position-momentum correlation measurement of photon pairs at telecommunication wavelength generated in a non-collinear, type-0 phase-matched SPDC process.
    For this purpose, we measured the position and momentum correlations of the generated signal and idler photons by recording the near- and far-field correlations, respectively, using a scanning technique.
    Based on the obtained correlation data, we certified EPR correlations, and consequently entanglement, of the signal-idler pairs with high statistical significance of $45$ standard deviations.
    We characterized the underlying high-dimensional entanglement further by estimating the entanglement of formation from our source.
    The calculated $E_F$ value greater than one implies an entanglement dimensionality of at least three in both scan directions.
    Note that the dimensionality can easily be increased by using a shorter crystal or a less focused pump-beam.
    To the best of our knowledge, this is the first experimental certification and characterization of position-momentum correlations from a type-0 phase-matched SPDC source at telecommunication wavelength.

    In a nutshell, we produced and certified position-momentum entanglement of photon pairs in the telecommunication wavelength regime which allows for the integration to telecommunication infrastructures and can be further extended to quantum hyper-correlations using several degrees of freedom in parallel.

\section*{Acknowledgments}
We gratefully acknowledge financial support from the Austrian Academy of Sciences and the EU project OpenQKD (Grant agreement ID: 85715).
E.A.O. and J.F. acknowledge ANID for the financial support (Becas de doctorado en el extranjero “Becas Chile”/2016 – No. 72170402 and 2015 – No. 72160487).
L.A. also thanks James Schneeloch and Marcus Huber for insightful comments.


\bibliography{biblio}
\bibliographystyle{apsrev4-1}

\end{document}